\documentclass[preprint,superscriptaddress,prd]{revtex4}
\usepackage{amssymb,amsmath,graphicx,amscd,xcolor,amsthm}

\theoremstyle{plain}

\def\E{\mathbf{E}}

\def\x{\mathbf{x}}

\def\v{\mathbf{v}}
\begin{document}

\title{Probing quantum vacuum fluctuations over a charged particle near a reflecting wall}
%
\author{V. A. \surname{De Lorenci}}
\email{delorenci@unifei.edu.br}
\affiliation{Instituto de F\'{\i}sica e Qu\'{\i}mica, Universidade Federal de Itajub\'a,\\
Itajub\'a, Minas Gerais 37500-903, Brazil}
\author{C. C. H. \surname{Ribeiro}}
\email{caiocesarribeiro@unifei.edu.br}
\affiliation{Instituto de F\'{\i}sica e Qu\'{\i}mica,    Universidade Federal de Itajub\'a,\\
Itajub\'a, Minas Gerais 37500-903, Brazil} 
\author{M. M. \surname{Silva}}
\email{silva.malumaira@gmail.com}
\affiliation{Dipartimento di Fisica, Universit\`a `La Sapienza',  I-00185 Roma, Italy}

\begin{abstract}
We investigate the influence of the vacuum fluctuations of a background electric field over a charged test particle in the presence of a perfectly reflecting flat wall. 
A  switching function connecting different stages of the system is implemented in such a way that its
functional dependence is determined by the ratio between the measuring time and the switching duration. 
The dispersions of the velocity components of the particle are found to be smooth functions of time, and have maximum magnitudes for a measuring time corresponding to about one round trip of a light signal between the particle and the wall. 
Typical divergences reported in the literature and linked with an oversimplification in modeling this system are naturally regularized in our approach.
Estimates suggest that this sort of manifestation of quantum vacuum fluctuations over the motion of the particle could be tested in laboratories. 
\end{abstract}

		
\maketitle

\section{Introduction}
A nonrelativistic test particle of charge $q$ and mass $m$,  placed at a distance $z$ from an infinite and perfectly reflecting flat wall, presents a stochastic motion induced by vacuum fluctuations of the electromagnetic field. This vacuum manifestation effect was reported \cite{ford2004,ford2005} to be characterized by the following dispersions (mean square deviations) of the particle velocity components in a time $\tau$:
\begin{equation}
\langle(\Delta v_{{}_\perp})^2\rangle = \frac{q^2}{\pi^2m^2}\frac{\tau}{32z^3}\ln\left(\frac{2z+\tau}{2z-\tau}\right)^2,
\label{larryvz}
\end{equation} 
\begin{equation}
\langle(\Delta v_{{}_\parallel})^2\rangle = \frac{q^2}{8\pi^2m^2}\left[\frac{\tau}{8z^3}\ln\left(\frac{2z+\tau}{2z-\tau}\right)^2
-\frac{\tau^2}{z^2(\tau^2-4z^2)}\right],
\label{larryvx} 
\end{equation} 
where the subscripts ``${}_\perp$" and ``${}_\parallel$" are hereafter used to indicate the dispersions of the velocity components perpendicular ($z$ direction)  and parallel ($x$ and $y$ directions) to the wall, respectively. Renormalization with respect to the background Minkowski vacuum state was implemented in these results. Unless otherwise explicitly stated, units are such that $\hbar=c=1$. Furthermore, we set the vacuum dielectric permittivity $\epsilon_0 \approx 8.85 \times 10^{-12}\, {\rm C^2/(N \,m^2)} =1$, which implies that $1V \approx 1.67\times 10^7 {\rm m}^{-1}$ in our units.

\begin{figure}[h!]
\includegraphics[width=8.5cm]{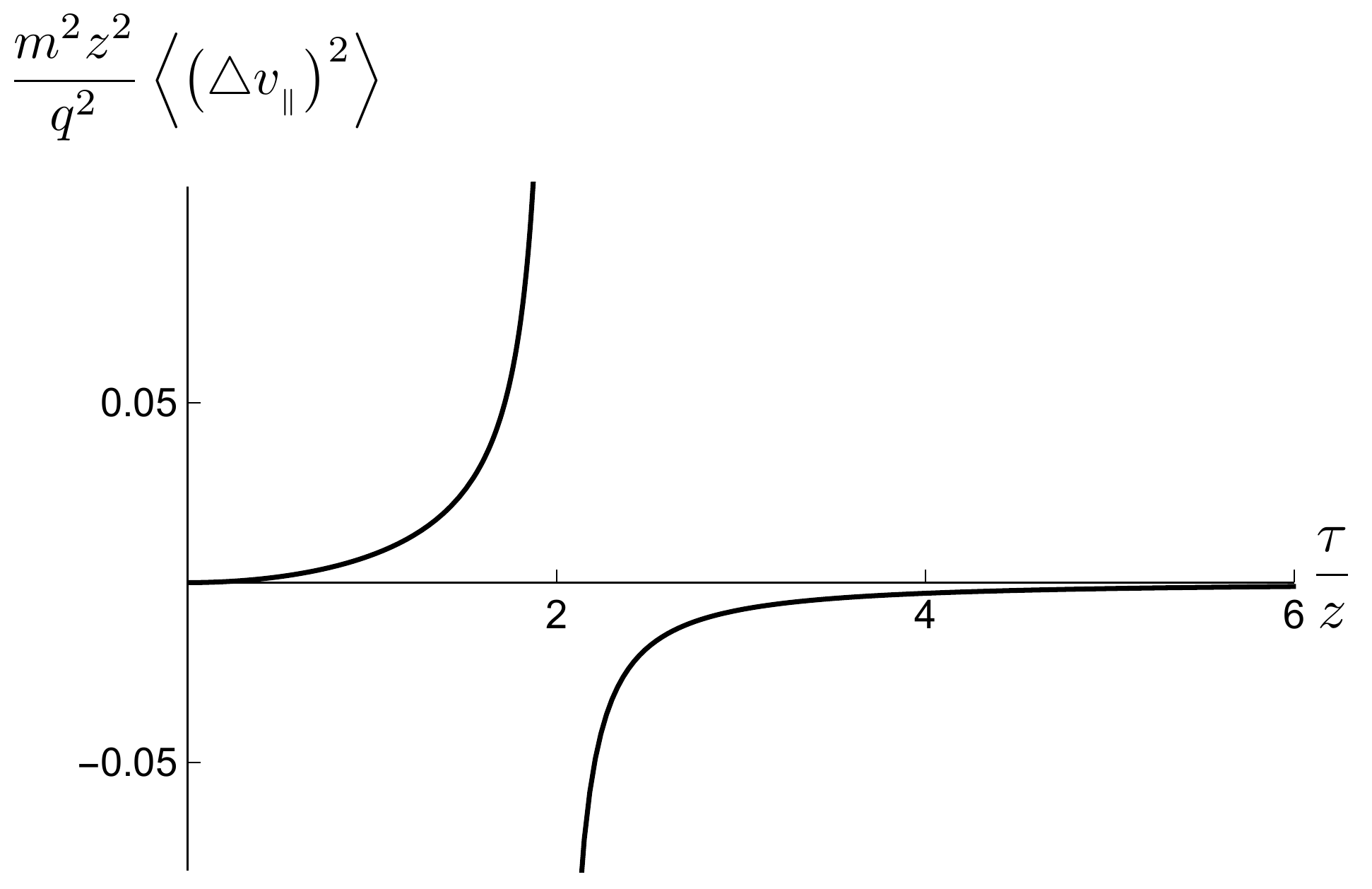}
\caption{Quantum dispersion of the parallel components of the particle velocity as a function of time.}
\label{fig-larry-x}
\end{figure}

\begin{figure}[h!]
\includegraphics[width=8.5cm]{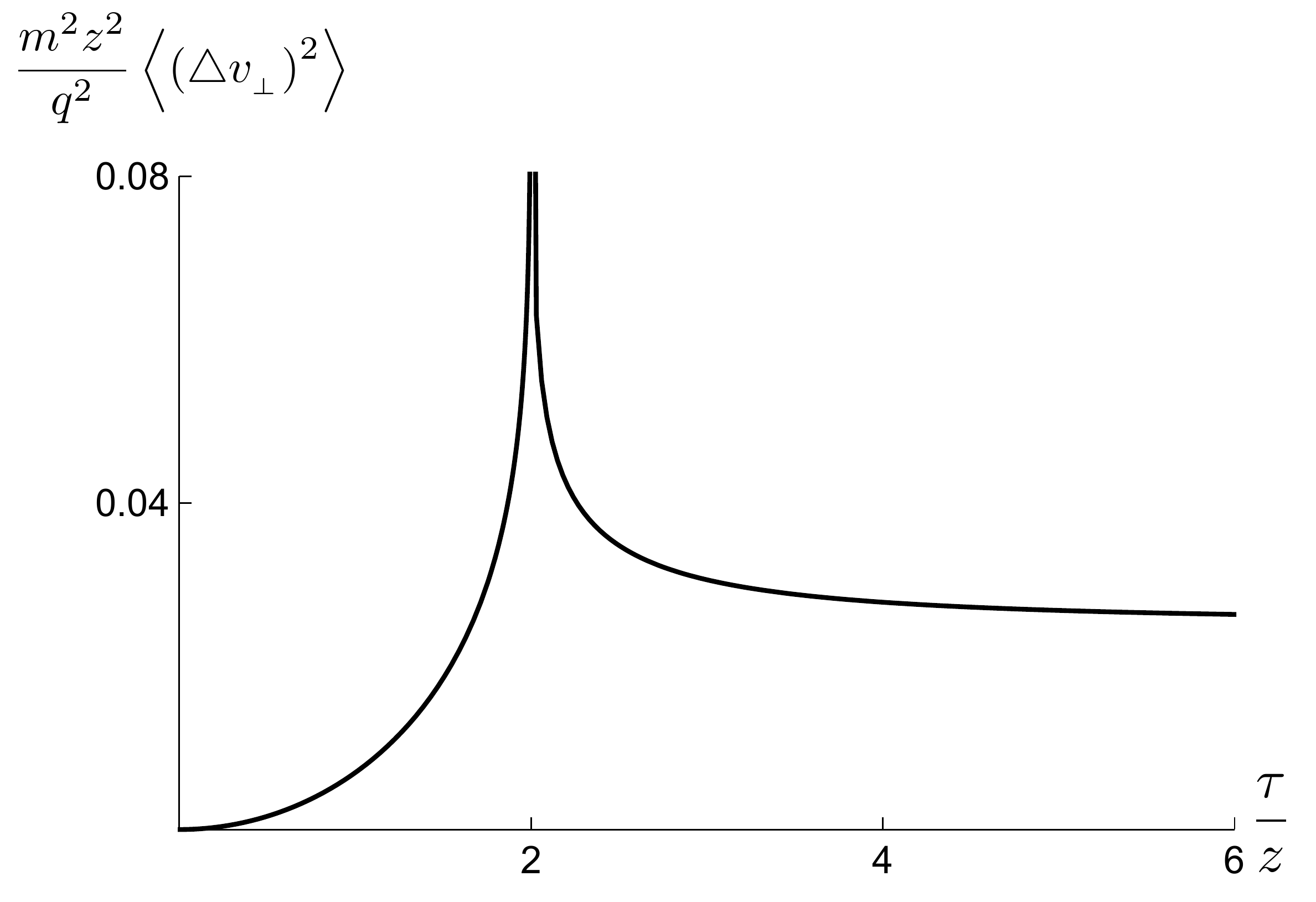}
\caption{Quantum dispersion of the perpendicular component of the particle velocity as a function of time.}
\label{fig-larry-z}
\end{figure}

Let us sum up the main features of the above results. 
The dispersion $\langle(\Delta v_{{}_\parallel})^2\rangle$ dies off as $\tau\rightarrow\infty$. However, in the perpendicular direction
$\langle(\Delta v_{{}_\perp})^2\rangle = q^2/(4 \pi^2m^2z^2)$ in this same limit,
suggesting a possible residual manifestation of the quantum vacuum fluctuations over the particle when a plane wall is present.
The behaviors of these dispersions as a function of time are depicted in Figs. \ref{fig-larry-x} and \ref{fig-larry-z}. 
This model was extended by including two plates \cite{hongwei2004} and finite-temperature effects \cite{hongwei2006}, and 
more recently a model for ``noncancellation of vacuum fluctuations'' based on this system was proposed  \cite{parkinson2011}.

Notice that $\langle(\Delta v_{{}_\perp})^2\rangle$ diverges at $z=0$, and both components  $\langle(\Delta v_{{}_\perp})^2\rangle$ and $\langle(\Delta v_{{}_\parallel})^2\rangle$ diverge at $\tau = 2z$, which corresponds to a round trip of a light signal between the test particle and the reflecting wall. 
It was suggested \cite{ford2004} that these divergences appear as a consequence of assuming an oversimplified model with a sharp boundary condition on the electric field at the perfectly reflecting wall.
Improved models were proposed by introducing a smooth (instead of a sudden) switching-on/off mechanism  \cite{seriu2008}, and later by considering a Gaussian wave-packet distribution in the time direction \cite{seriu2009}. In both cases some
different results were obtained, as compared to those in Ref. \cite{ford2004}. The main difference is the prediction that the perpendicular component of the velocity variance vanishes as $\tau\rightarrow\infty$, suggesting that no stationary behavior would remain. The focus of these works was on the temporal asymptotic regime. The results were not obtained exactly for all regimes. Furthermore, when the smooth switching mechanism was introduced \cite{seriu2008} only the dispersion of the perpendicular component of the velocity was examined.   
The existence of the above-mentioned divergences in a more realistic model has not been properly discussed so far. 
We notice however that when an idealized  Lorentzian switching is implemented \cite{seriu2008}, the divergence at $\tau=2z$ is naturally regularized. 
Nevertheless, the physical system studied here requires a switching function that assumes a nonzero constant value in a finite interval --- a flat measuring part \cite{seriu2008}. Such an interval characterizes the region where the particle is under the effect of the modified vacuum fluctuations induced by the presence of the wall. 
Despite the fact that a Lorentzian switching does not contain a flat measuring part (and thus it is not  suitable to model the system), this result already suggests that an appropriated switching mechanism would remove this divergence.

A similar system based on a quantum scalar field in (1+1) dimensions was recently studied \cite{delorenci2014}. In such a model it is concluded that allowing the particle position to fluctuate leads to a nonsingular behavior of the dispersion of its velocity.  An extension of this model for (3+1) dimensions has not yet been discussed. 

In this paper we go a step further in the analysis of the motion of a test charged particle driven by vacuum fluctuations in the presence of a reflecting wall. Following the idea introduced in Ref. \cite{seriu2008} we now implement a switching-on/off mechanism by means of a two-parameter sampling function connecting the two different stages of the system, i.e., the particle in empty space and the particle in the presence of a flat wall.  Exact analytic solutions for the dispersions in all components of the particle velocity are obtained for any point of space, at any time. No divergences appear. 
It is shown that after the particle is placed near the wall, the influence of the quantum vacuum fluctuations over it increases, achieving a maximum value at a finite time, and dies off as time goes to infinity.  No reminiscent effect survives. The behavior of such fluctuations around $\tau=2z$ is quite different for each direction of space. Parallel to the wall, the velocity dispersion achieves its maximum value just before $\tau=2z$, becomes negative just after that, and finally asymptotically goes to zero. On the other hand, in the perpendicular direction, the dispersion is always positive and achieves its maximum value  at about $\tau=2z$.

In the next section the system under study is presented in more detail, and the two-parameter switching function is introduced. The correlations of the background electric field modified by the presence of an idealized perfectly reflecting flat wall are given in Sec. \ref{correlations}. Dispersions of the velocity components of the test particle are derived in the following section. Different regimes are discussed, as well as the comparison with results previously reported in the literature. Some specific configurations are numerically studied. Estimates are presented in Sec. \ref{estimates}, where the layout of a possible experiment is proposed. It is suggested that the effects induced by vacuum fluctuations of the electric field over the trajectory of a charged test particle can be detected by means of a relatively simple experiment. Final remarks are presented in Sec. \ref{final}, and the derivation of the results presented in Sec. \ref{dispersions} are shown in full detail in the Appendix.

\section{The model}
\label{themodel}
The system under study is composed by a nonrelativistic test particle of mass $m$ and electric charge $q$ in the presence of a perfectly conducting flat wall. A background electric field $\E(\x,t)$ is prepared in its vacuum state (modified by the presence of the wall), for which $\langle \E(\x,t) \rangle = 0$, and we set the wall to be coincident with the plane $\{(x,y,z): z=0\}$. 

The interaction between the particle with the electric field is governed by the Lorentz force,
\begin{equation}
m\frac{d\v}{dt}  =  q \E(\x,t),
\label{1}
\end{equation}
where $\v= \v(\x,t)$.

The velocity of the particle in a given time $\tau$, which we identify as the measuring time, is found by integrating the equation of motion as
\begin{equation}
\v(\x,\tau)=\frac{q}{m}\int_{0}^\tau\E(\x,t)dt,
\label{eq1}
\end{equation}
where the particle was assumed to be at rest at the initial time $t_0 = 0\, {\rm s}$. 

Our analysis is restricted to the case where the particle displacement is negligible so that we may take its position $\x$ to be a constant \cite{ford2004}. 
Relaxing this condition amounts to assuming that $\x=\x(t)$, and time integrals involving $\x$ should be taken over the worldline of the particle. However, whenever low-velocity regime is assumed, it can be shown that corrections arising from this assumption are negligible. A specific case of  allowing the particle to be initially undergoing nonrelativistic motion in a given direction will be discussed in more detail later.

\subsection{Introducing a switching process} 
\label{secsw}
\begin{figure}[h!]
\includegraphics[width=8cm]{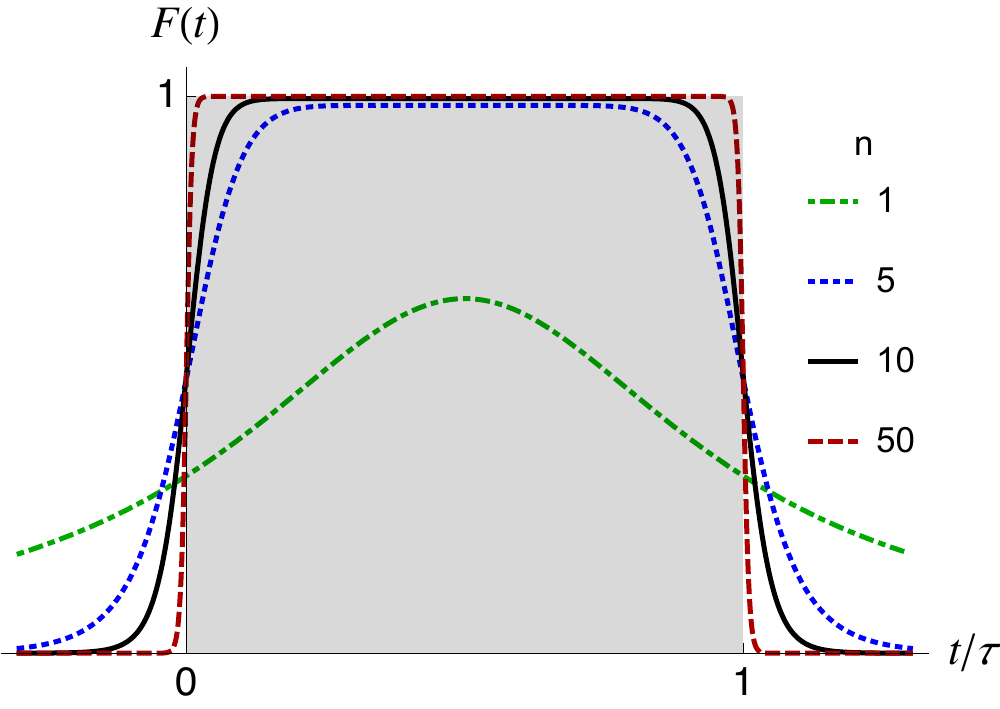}
\caption{Smooth curves given by Eq. \ref{sample} (with a dislocated origin) in contrast with the Heaviside-like behavior, represented by the background rectangle.}
\label{fig1}
\end{figure}

Equation (\ref{eq1}) admits a sort of sudden switching process, where the interaction is instantaneously turned on at an initial time $t_0=0 \,{\rm s}$ and turned off after an interval of time $\tau$. In a more realistic setup,  finite intervals of time associated with starting (switching on) and ending (switching off) an interaction regime should be considered. 
Generically, the switching-on/off operations are implemented by introducing a sample function $F(t)$ in the integration appearing in Eq. (\ref{eq1}) as
\begin{equation}
\int_{0}^\tau\E(\x,t)dt \rightarrow \int_{-\infty}^\infty\E(\x,t)F(t)dt,
\label{switching}
\end{equation}
with $F(t)$ presenting a behavior like any of those exhibited by the curves in Fig. \ref{fig1}.
Notice that Eq. (\ref{eq1}) is just a particular case of this equation, corresponding to a sudden measurement process, for which the switching function is identified with a product of Heaviside functions as $F(t) \rightarrow \Theta(t)\Theta(\tau-t)$. 
By implementing Eq. (\ref{switching}) we are assuming that a smooth transition must exist between the two different stages: a particle near the wall and a particle in empty space. For instance, we can imagine that the wall is placed near the particle in a certain interval of time (the switching time), after which it is removed, or that the particle moves from an empty region of space to the region where the wall is placed. In both cases the interaction between the particle and the wall effectively begins before $t=0\,s$ and still exists after $t=\tau$. This aspect is clearly satisfied when a smooth transition is assumed. The switching duration will depend on how fast the system is set up. Hence, the choice of the switching function depends on the details of the system. 

The use of smooth sample functions to define well-behaved operators in quantum field theory has been been considered in the literature for quite some \cite{streater1964}. The implementation of this technique in the description of physical systems appears in some cases \cite{seriu2008,seriu2009}, including the study of light propagation in nonlinear medium under the influence of background quantum fluctuations of an external electric field (see Ref.  \cite{bessa2016} and references therein). 

We shall assume in our model a switching process governed by one of the functions in the sequence of functions
\begin{equation}
F(t) \rightarrow F^{(n)}_\tau(t) = \frac{c_n}{1+\left(\frac{2t}{\tau}\right)^{2n}},
\label{sample}
\end{equation}
where $n$ is a positive integer, and $c_n = (2n/\pi)\sin(\pi/2n)$ is the normalization constant, which is taken such that
\begin{equation}
\int_{-\infty}^\infty F^{(n)}_\tau(t)dt = \tau.
\end{equation}
The curves determined by these functions, which are characterized by the parameter $n$ and the time interval $\tau$, are depicted in Fig. \ref{fig1} for several values of $n$. Notice that as $n$ gets larger, the corresponding function approaches the Heaviside-like behavior, pictorially represented by the background rectangle, while for $n=1$ it recovers the Lorentzian distribution. 

The values of $n$ and $\tau$ for a given system may be constrained by the experimental setup. Specifically, the switching interval of time, here defined as $\tau_{{}_S}$, determines the value of $n$, and it can be calculated by taking the difference between the nearest points of maximum curvature of $F^{(n)}_\tau(t)$, which are given by the zeros of the third derivative of $F^{(n)}_\tau(t)$. Straightforward calculation leads to
\begin{eqnarray}
\tau_{{}_S} =&& \frac{\tau}{2}\left(\frac{2n-1}{n+1}\right)^{\frac{1}{2n}}\left[\left(1+\sqrt{1-\frac{(n+1)(n-1)}{(2n+1)(2n-1)}}\,\right)^{\frac{1}{2n}} \right.
\nonumber\\
&&\left.-\left(1-\sqrt{1-\frac{(n+1)(n-1)}{(2n+1)(2n-1)}}\,\right)^{\frac{1}{2n}}\right].
\label{ts}
\end{eqnarray}
A sudden process consists in an idealization for which $\tau_{{}_S} \rightarrow 0$, and corresponds to taking $n\rightarrow\infty$. 

Switching functions with a finite duration \cite{fewster2015} are also useful to describe smooth transitions in physical systems, such as the one we are here examining.  However, regarding the magnitude of the effects induced by vacuum fluctuations, for large values of $n$ both methods may lead to the same results, as contributions related to the tails will be completely negligible.

\section{Correlations of the electric field}
\label{correlations}
We wish to study the case where the charged particle is subjected to a fluctuating electric field $\E(\x,t)$. More specifically, in our approach the electric field $\E(\x,t)$ is a purely quantum quantity and has a null mean value $\langle \E(\x,t)\rangle$, but presents non-zero second-order correlations $\langle \E(\x,t)\E(\x',t') \rangle$. The presence of correlations of the electric field induces a sort of stochastic-motion effect on the particle, the consequences of which will be investigated in the following sections. Once $\langle \E(\x,t)\rangle=0$, it follows that $\langle \v\rangle = 0$.
Hence, an important quantity to be determined is the dispersion of the particle velocity $\langle (\Delta v_i)^2\rangle = \langle v_i^2 \rangle - \langle v_i \rangle^2$.

We notice that, in the absence of thermal effects \cite{hongwei2006}, if $\E(\x,t)$ also has a classical part, only its quantum contributions would take part in the result for the dispersion of $v_i$. Classical terms in $\langle v_i^2 \rangle$ would identically cancel out with those terms coming from $\langle v_i \rangle^2$, in the computation  of $\langle (\Delta v_i)^2 \rangle$ \cite{ford2004}. 

The needed electric field correlation functions in the presence of a perfectly reflecting wall are given by \cite{brown1969} 
\begin{align}
&\langle E_x(\x,t)E_x(\x',t')\rangle 
\nonumber\\
&= \frac{1}{\pi^2}\left[\frac{(\Delta t)^2-(\Delta x)^2+(\Delta y)^2+(\Delta z)^2}{\left[(\Delta t)^2-(\Delta \x)^2\right]^3}\right]
\nonumber\\
&-\frac{1}{\pi^2}\left[\frac{(\Delta t)^2-(\Delta x)^2+(\Delta y)^2+(z+z')^2}{\left[(\Delta t)^2 -(\Delta x)^2-(\Delta y)^2-(z+z')^2\right]^3}\right],
\label{cx}
\end{align} 
\begin{align}
&\langle E_y(\x,t)E_y(\x',t')\rangle 
\nonumber\\
&= \frac{1}{\pi^2}\left[\frac{(\Delta t)^2+(\Delta x)^2-(\Delta y)^2+(\Delta z)^2}{\left[(\Delta t)^2-(\Delta \x)^2\right]^3}\right]
\nonumber\\
&-\frac{1}{\pi^2}\left[\frac{(\Delta t)^2+(\Delta x)^2-(\Delta y)^2+(z+z')^2}{\left[(\Delta t)^2 -(\Delta x)^2-(\Delta y)^2-(z+z')^2\right]^3}\right],
\label{cy}
\end{align} 
\begin{align}
&\langle E_z(\x,t)E_z(\x',t')\rangle 
\nonumber\\
&= \frac{1}{\pi^2}\left[\frac{(\Delta t)^2+(\Delta x)^2+(\Delta y)^2-(\Delta z)^2}{\left[(\Delta t)^2-(\Delta \x)^2\right]^3}\right]
\nonumber\\
&+\frac{1}{\pi^2}\left[\frac{(\Delta t)^2+(\Delta x)^2+(\Delta y)^2-(z+z')^2}{\left[(\Delta t)^2 -(\Delta x)^2-(\Delta y)^2-(z+z')^2\right]^3}\right],
\label{cz}
\end{align}
where $\Delta$ is a symbolic operator whose action on a quantity $a$ is defined by $\Delta a \equiv a-a'$. 
The first term in each of the above correlations corresponds to the Minkowski vacuum contribution, and the second one is due to the presence of the wall.

\section{Dispersions}
\label{dispersions}
Now suppose the particle is placed at a distance $z$ from the flat wall in a given initial time. The presence of the wall will modify the Minkowski vacuum fluctuations of the electromagnetic field and will produce an effect on the state of motion of the particle that can be measured by looking at the fluctuations of its velocity. 
As previously discussed, after introducing the switching function, the velocity of the particle in a given time $\tau$ will be given by
\begin{equation}
\v(\x,\tau)=\frac{q}{m}\int_{-\infty}^{\infty} F^{(n)}_\tau(t) \E(\x,t)dt.
\label{eq1b}
\end{equation}
As the background field has a null mean value in the modified vacuum state ($\langle \E\rangle =0$) we obtain that $\langle(\Delta v_i)^2\rangle = \langle v_i^2\rangle - \langle v_i\rangle^2 = \langle v_i^2\rangle$. Hence,
\begin{eqnarray}
\langle(\Delta v_i)^2\rangle =  \frac{q^2}{m^2}\int_{-\infty}^\infty\int_{-\infty}^\infty \langle E_i(\x,t)E_i(\x,t')\rangle_{{}_{\tt Ren}} \nonumber\\
\times F^{(n)}_\tau(t)F^{(n)}_\tau(t') dt dt'.
\label{eq2}
\end{eqnarray}
Here, the renormalization ({\tt Ren}) procedure was implemented by subtracting the Minkowski vacuum contribution from the two-point function $\langle E_i(\x,t)E_i(\x,t')\rangle$, which means dropping the terms in Eqs. (\ref{cx}),  (\ref{cy}), and (\ref{cz}) that diverge when the limit of point coincidence is taken.

Thus, taking the limit $\x'\rightarrow \x$ in the renormalized versions of Eqs. (\ref{cx}),  (\ref{cy}) and  (\ref{cz}),
\begin{eqnarray}
\langle E_x(\x,t)E_x(\x,t')\rangle_{{}_{\tt Ren}} &=& \langle E_y(\x,t)E_y(\x,t')\rangle_{{}_{\tt Ren}} 
\nonumber \\
&=& \frac{(\Delta t)^2 +4z^2}{\pi^2[4z^2-(\Delta t)^2]^3},
\label{c1}
\\
\langle E_z(\x,t)E_z(\x,t')\rangle_{{}_{\tt Ren}} &=&\frac{1}{\pi^2[4z^2-(\Delta t)^2]^2},
\label{c2}
\end{eqnarray}
and inserting these results [together with Eq. (\ref{sample})] into Eq. (\ref{eq2}), we obtain
\begin{eqnarray}
\langle(\Delta v_{{}_\parallel})^2\rangle = \frac{q^2c_n^2}{\pi^2m^2}\int_{-\infty}^\infty\int_{-\infty}^\infty 
\frac{(t-t')^2 +4z^2}{\big[4z^2-(t-t')^2\big]^3}
\nonumber\\
\times\frac{1}{\left[1+\left(\frac{2t}{\tau}\right)^{2n}\right]\left[1+\left(\frac{2t'}{\tau}\right)^{2n}\right]} dt dt'
\label{eqvx2}
\\
\langle(\Delta v_{{}_\perp})^2\rangle = \frac{q^2c_n^2}{\pi^2m^2}\int_{-\infty}^\infty\int_{-\infty}^\infty 
\frac{1}{\big[4z^2-(t-t')^2\big]^2}
\nonumber\\
\times\frac{1}{\left[1+\left(\frac{2t}{\tau}\right)^{2n}\right]\left[1+\left(\frac{2t'}{\tau}\right)^{2n}\right]} dt dt'.
\label{eqvz2}
\end{eqnarray}
We recall that $\langle(\Delta v_x)^2\rangle=\langle(\Delta v_y)^2\rangle \doteq \langle(\Delta v_{{}_\parallel})^2\rangle$.

The integrals appearing in the above expressions can be analytically solved (see Appendix for full detail), resulting in
\begin{eqnarray}
\langle(\Delta v_{{}_\parallel})^2\rangle = - \left(\frac{2qc_n}{mn\tau}\right)^2\;\sum_{p=0}^{n-1}\sum_{q=n}^{2n-1}
\psi_{n,p}\psi_{n,q}
\nonumber\\
\times\frac{\left(\psi_{n,p}-\psi_{n,q}\right)^2+(4z/\tau)^2}{\left[\left(\psi_{n,p}-\psi_{n,q}\right)^2-(4z/\tau)^2\right]^3},
 \label{53}
 \\
 \langle(\Delta v_{{}_\perp})^2\rangle = \left(\frac{2qc_n}{mn\tau}\right)^2\;\sum_{p=0}^{n-1}\sum_{q=n}^{2n-1}
\psi_{n,p}\psi_{n,q}
\nonumber\\
\times
\frac{1}{\left[\left(\psi_{n,p}-\psi_{n,q}\right)^2-(4z/\tau)^2\right]^2},
 \label{53b}
\end{eqnarray}
where we have defined $\psi_{n,s} \doteq \exp[{i(\pi/2n)(1+2s)}]$, for any $s$.

\begin{figure}[h!]
\includegraphics[width=8.7cm]{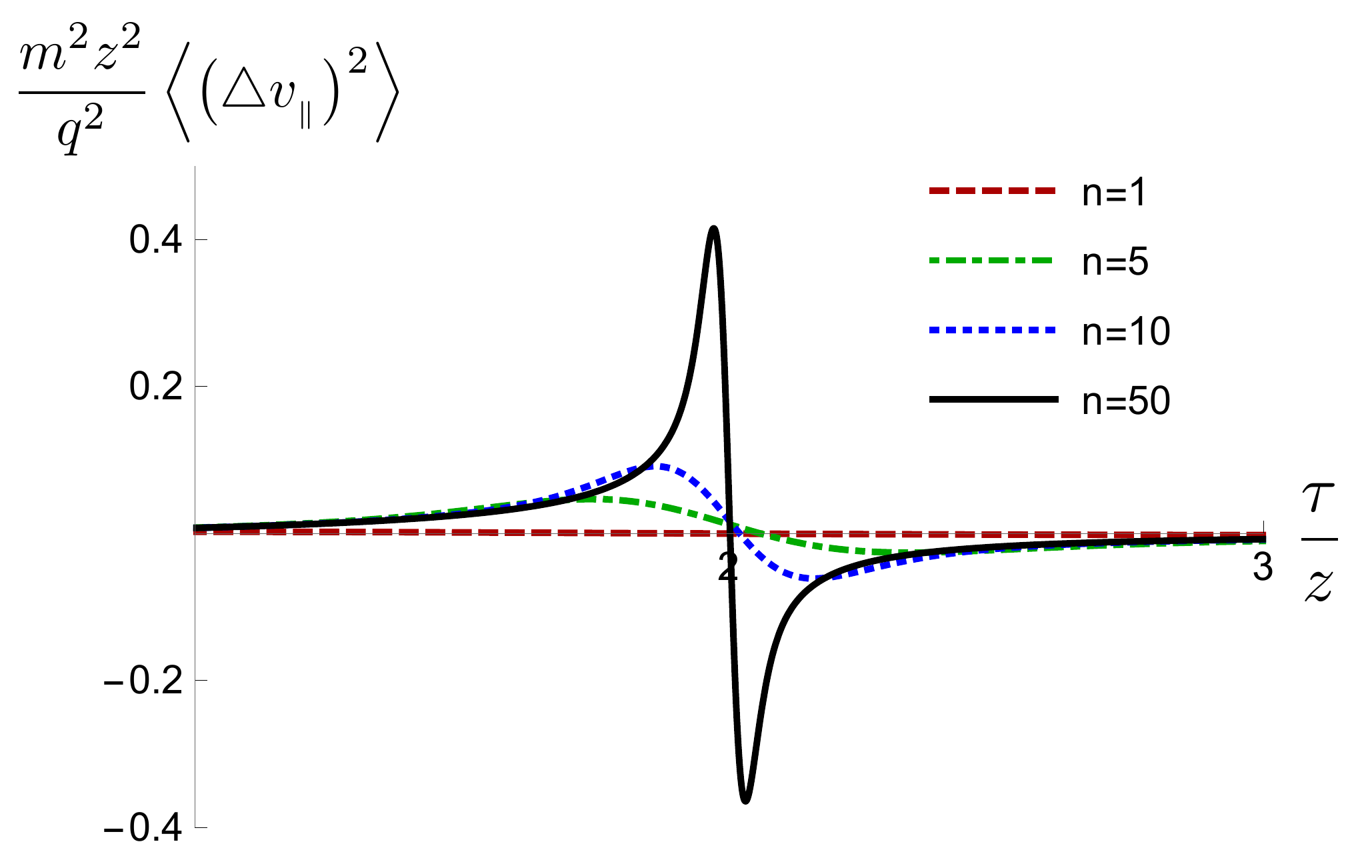}
\caption{Dispersion of the parallel components of the particle velocity as a function of $\tau$ for several values of the parameter $n$. Notice that the dispersion becomes negative after $\tau=2z$.}
\label{fig-variance-x}
\end{figure}

\begin{figure}[h!]
\includegraphics[width=8.7cm]{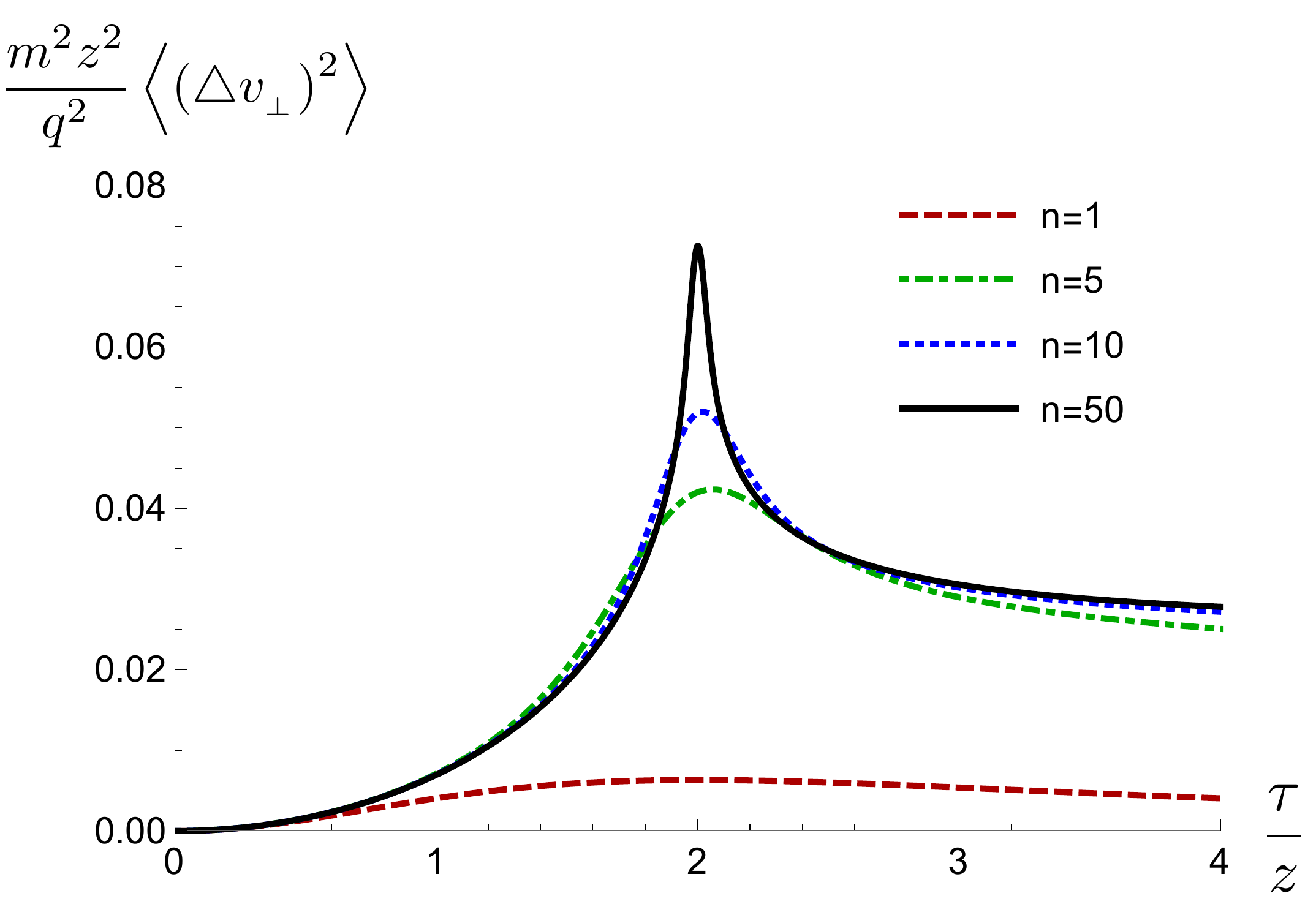}
\caption{Dispersion of the component of the particle velocity perpendicular to the wall as a function of $\tau$ for several values of the parameter $n$. Notice that the maximum effect occurs just about $\tau=2z$.}
\label{fig-variance-z}
\end{figure}

The behaviors of these dispersions as function of $\tau$ for several values of $n$ are depicted in Figs. \ref{fig-variance-x} and \ref{fig-variance-z}.   
One first aspect one should notice is that they are well behaved at $\tau = 2z$. The divergences appearing in former investigations \cite{ford2004,hongwei2004,ford2005} are thus naturally regularized in our model. As pointed out \cite{ford2004}, $\tau = 2z$ corresponds to the duration of a round trip of a light signal between the charged particle and the flat wall. Figure \ref{fig-variance-x} shows that $\langle(\Delta v_{{}_\parallel})^2\rangle$ is nearly zero at $\tau=2z$, for any value of $n$. However,  it achieves its maximum magnitude (positive or negative) just around this specific time. Another point to be noticed is that this dispersion becomes negative after $\tau = 2z$. Negative dispersions appear as a consequence of the renormalization with respect to the Minkowski vacuum state. This means that, in this case, the presence of the wall decreases the always-present background Minkowski vacuum fluctuations. 
On the other hand, as depicted in Fig. \ref{fig-variance-z}, the dispersion  $\langle(\Delta v_{{}_\perp})^2\rangle$ is  always a positive function of time, and achieves its maximum value around $\tau = 2z$, for any value of $n$. We stress that the value of $n$ is associated with the switching time $\tau_{{}_S}$, and thus depends on experimental arrangements.
As both figures show, the dispersions are initially zero, present a transient behavior around $\tau=2z$, and evolve asymptotically to zero as $\tau\rightarrow\infty$. In fact, Eqs. (\ref{53}) and (\ref{53b}) reveal that $\langle(\Delta v_i)^2\rangle \sim 1/\tau^2$ in this asymptotic regime. These results clearly show that no reminiscent effect occurs in this system, and the behavior of the dispersions is closely related with the way the system is configured. 
Finally, for any $\tau$ the dispersions are finite as we approach the wall at $z=0$, and go to zero when $\tau$ goes to zero, as expected.

Before closing this section, let us look at a scenario in which the particle is assumed to be initially in nonrelativistic motion with constant velocity $v_{{}_0}$, say in $x$ direction. Now the vacuum expectation value of the particle velocity would be different from zero, i.e., $\langle v_i \rangle = v_{{}_0}\delta_{ix}$. Besides this contribution, which is exactly canceled out in the calculations leading to the dispersion $\langle(\Delta v_i)^2\rangle$, only corrections of the order ${\cal O}(v_{{}_0}^2)$ would appear in the results presented by Eqs. (\ref{53}) and (\ref{53b}). For instance, it can be shown that the dispersion in the perpendicular direction, which is of particular interest to us, would be slightly modified as
\begin{eqnarray}
\langle(\Delta v_{{}_\perp})^2\rangle \approx \left(\frac{2qc_n}{mn\tau}\right)^2\;\sum_{p=0}^{n-1}\sum_{q=n}^{2n-1}
\psi_{n,p}\psi_{n,q}
\nonumber\\
\times
\frac{\left(\psi_{n,p}-\psi_{n,q}\right)^2(1+4v_{{}_0}^2)-(4z/\tau)^2}{\left[\left(\psi_{n,p}-\psi_{n,q}\right)^2-(4z/\tau)^2\right]^3}.
 \label{53bb}
\end{eqnarray}
Thus, whenever such corrections are negligible, our results can be safely applicable when the particle is assumed to be initially in nonrelativistic constant motion.

\section{Estimates}
\label{estimates}
As seen in Sec. \ref{secsw}, the parameter $n$ introduced by $F^{(n)}_\tau(t)$ depends upon the switching interval of time $\tau_{{}_S}$. In order to examine a specific example, suppose the system is prepared in such a way that a charged particle, initially in an empty region, experiences the presence of a perfectly reflecting plane wall of size $L$ during a time $\tau$. Assume the transition between the different stages (the empty region and the region with the wall) occurs in an interval of time corresponding to one-hundredth of $\tau$. This means that $\tau_{{}_S}/\tau = 0.01$, and it corresponds to choosing the sample function $F^{(n)}_\tau(t)$ with $n=65$, as given by Eq. (\ref{ts}).  
As a possible realization of the switching \cite{ford2004}, suppose the particle is an electron (charge $q_e$ and mass $m_e$), and that it has a constant initial velocity $v_{{}_0}$ in the $x$ direction, which is parallel to the plane wall. 
It comes from an empty region, passes by the wall, and continues traveling through empty space. 
The travel time of the particle across the wall is thus $L/v_{{}_0}$, and corresponds approximately to the time interval during which the electron is under the influence of the modified vacuum fluctuations due to the presence of the wall. This interval of time is thus identified with the measuring time $\tau$. Hence, $\tau_{{}_S} \approx 0.01 L/v_{{}_0}$. Reintroducing dimensionful $c$, we conveniently write
\begin{equation}
(z/L) = (z/c\tau)(c/v_{{}_0}).
\label{just}
\end{equation}
Our formalism is applicable provided that $z/L <<1$ and $v_{{}_0}/c <<1$, as discussed in the previous section. 

As Fig. \ref{fig-variance-z} shows, the maximum effect produced by the vacuum fluctuations occurs just about a measuring time $\tau=2z/c$. However, as long as a not too large wall is considered, the use of Eq. (\ref{just}) shows that $\tau=2z/c$ is out of experimental probing. 

\begin{figure}[h!] 
\includegraphics[width=8.7cm]{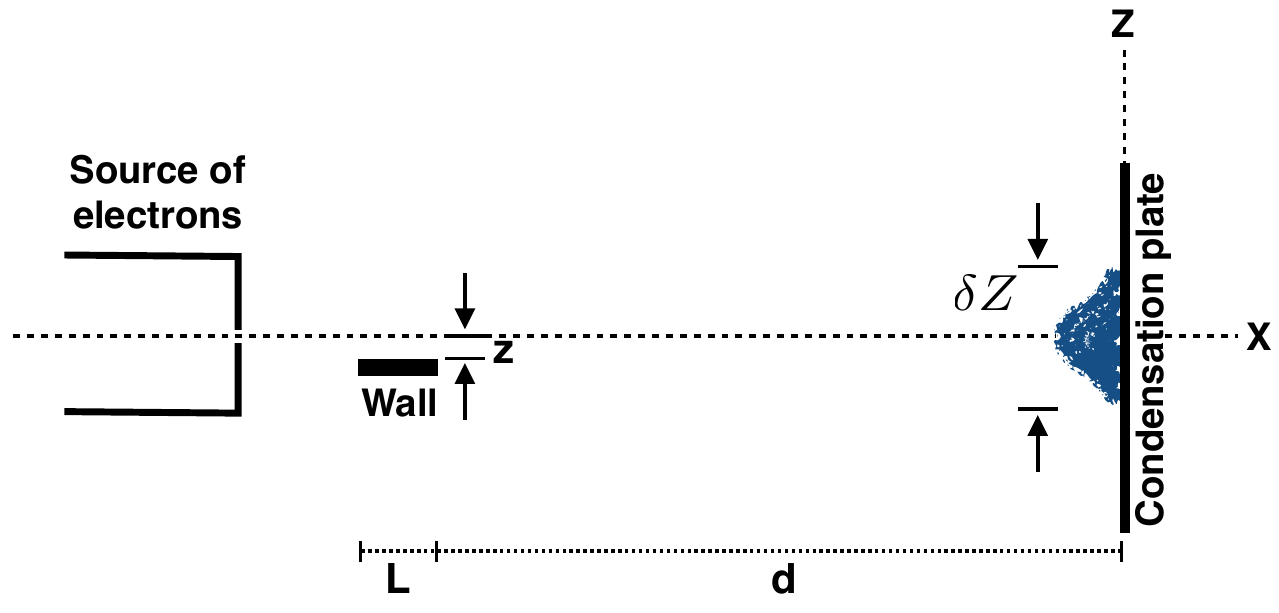}
\caption{Layout (out of scale) of a possible experiment designed to measure the dispersions of the particle velocity induced by vacuum fluctuations. The particle is emitted by a source in the $x$ direction, has its motion influenced by the modified quantum vacuum fluctuations in the wall region, and is measured in a condensation plate.}
\label{exp}
\end{figure}

 A possible layout for an experiment is shown in the Fig. \ref{exp}. 
The values of the magnitude of the dispersion of the perpendicular component of the particle velocity  for suggestive values of the measuring time $\tau$ are listed in Table \ref{tab2}.
\begin{table}[h!]
   \centering
    \caption{Some values for the magnitude of $\sqrt{\langle(\Delta v_{{}_\perp})^2\rangle}$ as a function of $\tau$. Here we assume $n=65$ and that the test particle is an electron.}
   \begin{tabular}{lccccc } 
   \hline\hline
      $c\tau/z$                       & 2      &50       &100     &500     &1000   \\
      \hline
       $(z/1{\rm m})(\sqrt{\langle(\Delta v_{{}_\perp})^2\rangle}/c)\times 10^{14}$ &$3.22$&$1.80$&$1.56$&$0.47$&$0.24$\\
       \hline\hline
          \end{tabular}
   \label{tab2}
\end{table}

Let us select the case where $\tau = 1000z/c$, for which
 \begin{equation}
 \sqrt{\langle(\Delta v_{{}_\perp})^2\rangle} \approx 0.7 \left(\frac{m_e}{m}\right)\left(\frac{q}{q_{e}}\right) \left(\frac{1 \mu{\rm m}}{z}\right) {\rm m\,s^{-1}}.
 \label{exp-v}
 \end{equation}
After the particle leaves the wall region, it will have gained an uncertainty in its velocity induced by the vacuum fluctuations whose magnitude in the $i$th direction is given by $\sqrt{\langle(\Delta v_{i})^2\rangle}$. The interval of time in which the electron moves from the wall to the condensation plate is given by $t_d \approx d/v_{{}_0}$. 
Hence, concentrating on the effect in the $z$ direction, an electron is expected to arrive at the condensation plate within a range of positions given by  2$\sqrt{\langle (\Delta v_{{}_\perp})^2\rangle}\; t_d = 2\sqrt{\langle (\Delta v_{{}_\perp})^2\rangle}\; d/v_{{}_0} \doteq \delta Z$. 
A possible realization of the above experiment could be achieved by setting  $v_{{}_0}/c = 0.05$, $z/L = 1/50$, and $\tau=1000z/c$. With this configuration, we would obtain
\begin{equation}
\delta Z = 9.4 \left(\frac{m_e}{m}\right)\left(\frac{q}{q_{e}}\right) \left(\frac{1\mu {\rm m}}{z}\right)\left(\frac{d}{100 {\rm m}}\right)\mu {\rm m}.
 \label{exp-z}
\end{equation}
This seems to be a measurable consequence of the quantum fluctuations of the electromagnetic field.  Notice that the result can be improved by setting a longer distance $d$. For instance, by setting $d=1{\rm km}$ we would obtain  $\delta Z \approx 0.1 {\rm mm}$.

The results presented in Eqs. (\ref{exp-v}) and (\ref{exp-z}) are generically expressed in terms of distances $z$ and $d$ and ratios $m/m_e$ and $q/q_e$, where $m$ and $q$ are mass and charge of any test particle, respectively. Hence, it is a straightforward numerical exercise to obtain the magnitude of the effect for heavier particles in different configurations, provided the limits of applicability of the model are observed.  

The speed of the electron in the above estimates is just $0.05c$, which leads to a relativistic factor of the order $\gamma \approx 1.001$. In this regime we do not expect to have contributions from the dynamical Casimir effect. However, as pointed out in Ref. \cite{ford2004}, radiation is expected to be emitted by the electron as a switching effect. Considering the specific model studied in the last section, in which the electron moves parallel to the wall, the angular spectral density of radiation \cite{karlovets2008} from the electron has a magnitude that decreases exponentially with the distance from the electron to the edge of the wall. Just to have an estimate, it can be shown that the total energy emitted perpendicularly to the wall (in the entire frequency range) when the electron is in its maximum approximation to the edge of the wall is approximately given by $W \approx v_{{}_0}{q_e}^2/8\pi^2 c z  \approx 1.145\times 10^{-5}{\rm eV}$. Now, if we assume that this energy is concentrated in a single photon it would lie in the microwave frequency band, and its backreaction over the electron velocity would be of the order of $10^{-2} {\rm m\,s^{-1}}$. Hence, this effect also contributes to the total dispersion of the electron velocity, but with a lower magnitude. 
The contribution of this effect to the total dispersion of the particle velocity could be tracked by measuring the radiation emitted by the electron during the switching, and it could eventually be experimentally distinguished from the contribution produced by vacuum fluctuations. 

\section{final remarks}
\label{final}

In a previous work \cite{delorenci2014} a simplified (1+1)-dimensional model based on a test particle interacting with a quantum scalar field near a perfectly reflecting boundary was investigated. It was shown that the dispersion of the velocity of the test particle exhibits a behavior similar to that of a charged particle under the influence of vacuum fluctuations of the electric field in the presence of a reflecting boundary \cite{ford2004}. 
In such a simplified model, when a Gaussian blurring is performed on the position of the particle (which partially models its quantum behavior) it is shown that the divergence at $\tau=2z$ is regularized. However, when generalized to higher dimensions, it can be shown that this kind of regularization does not work on all components of the particle velocity.  
We suspect that this method of introducing a blurring in the particle position will work well for a particle in a box, because then there would be boundary conditions in all spatial directions. This issue deserves further investigation.

A new aspect introduced in this paper was the implementation of a smooth two-parameter switching function connecting the distinct stages in the description of the system. As a consequence, the dispersions in all components of the particle velocity were found to be non-singular functions of time and distance to the wall. Divergences previously found in the literature, and linked to an oversimplification of the model, were naturally regularized. In particular, we conclude that the divergence at $\tau=2z$ is more closely related with the assumption of a sudden switching approach than that of a sharp boundary condition on the wall.
Additionally, the switching functions we introduced have the advantage of allowing analytic integration of the dispersions, which gave handle results that can be easily adapted to possible experimental situations.
Our results suggest that the quantum-vacuum-induced effects on the motion of a charged test particle could be tested by means of measuring the dispersions of the trajectory of the particle.
Another important result we found is the late-time behavior of the dispersions when the smooth switching is implemented. Compared to the sudden switching [which leads to a finite dispersion $\langle(\Delta v_{{}_\perp})^2\rangle$ even in the limit $\tau/z \rightarrow \infty$], when the smooth switching is implemented the dispersions of all velocity components go away in this limit, which is expected from the energy conservation law. The dispersions appear as consequence of the way the system is set up, which is translated in terms of the specific switching that is implemented.

A sharp boundary condition and sudden switching are rather convenient idealized assumptions that keep calculations simple, but they usually lead to singular behavior of certain physical quantities. There are several different methods of implementing regularization in problems involving idealized boundaries. The central idea is to perform a blur of the boundary position.
A different way to look for a regularization of the above-mentioned divergences would be by assuming more realistic boundaries, such as considering the particle near a half-space filled with a dielectric material. Quantum correlations of electric and magnetic fields near a homogenous nondissipative  dielectric have already been considered in the literature (see, for instance, \cite{ford2002,passante2012}). It would be interesting to investigate the effect studied here in the presence of such boundaries.

The existence of a classical interaction between the charged particle and the conducting wall was ignored in our study. 
It is a well-known fact that a charged particle interacts with a conducting wall and is affected by a force caused by the redistribution of free charge on the wall. However, such an effect is only responsible for a deflection of the trajectory of the test particle, with no contributions to quantum dispersions. For instance, in the figure illustrating a possible experiment to measure the consequences of such dispersions, the same distribution of arrival positions of the electron around a central point would be observed, but the whole distribution would appear shifted by a certain distance, say  $z_{{}_0}$, caused by the classical effect. 

As a final remark, in Sec. IV we implemented renormalization with respect to the Minkowski vacuum, i.e., we subtracted from the dispersions the terms related to the free-space electric field correlation functions. So, the modified vacuum is solely responsible for the effect we studied. 
In fact,  free-space vacuum fluctuations do not induce measurable effects on the motion of the test particle. This is because such fluctuations are highly anticorrelated \cite{parkinson2011}.  
As a concrete example, Ref. \cite{johnson2002} explicitly showed that test particles in the scalar-field free vacuum do not experience any influence on their motion. 
Furthermore, we should point out that in the absence of a boundary there is no sense of using a switching. 

\begin{acknowledgments}
We would like to thank L.H. Ford, E.S. Moreira Jr. and R. Rischter for useful discussions. This work was partially supported by the Brazilian research agencies CNPq (Conselho Nacional de Desenvolvimento Cient\'{\i}fico e Tecnol\'ogico) under Grant No. 302248/2015-3, FAPEMIG (Funda\c{c}\~ao de Amparo \`a Pesquisa do Estado de Minas Gerais), and CAPES (Coordena\c{c}\~ao de Aperfei\c{c}oamento de Pessoal de N\'{\i}vel Superior).
\end{acknowledgments}

\section*{Appendix: Analytic Solution of the Integrals}
Here we show in detail the technique used to solve the integrations in Eqs. (\ref{eqvx2}) and (\ref{eqvz2}). We define the classes of integrals $I_{\tau,ijkl}^{(n)}$ given by
\begin{eqnarray}
I_{\tau,ijkl}^{(n)}=\int_{-\infty}^\infty\int_{-\infty}^\infty\frac{(t-t')^i +(2z)^j}{\left[(2z)^k-(t-t')^2\right]^l} 
\nonumber\\
\times\frac{1}{\left[1+\left(\frac{2t}{\tau}\right)^{2n}\right]\left[1+\left(\frac{2t'}{\tau}\right)^{2n}\right]}dtdt',
\end{eqnarray}
where $n$, $i$, and $l$ are positive integers, $k$ and $j$ are real numbers, and the condition $i < 2l$ is observed. Implementing the change of variables $u=2t/\tau$, $u' = 2t'/\tau$, and defining $a \doteq (2z)^j(2/\tau)^i$ and $b \doteq (2z)^k(2/\tau)^2$, it follows that
\begin{eqnarray}
I_{\tau,ijkl}^{(n)}=\left(\frac{\tau}{2}\right)^{2+i-2l}
\int_{-\infty}^\infty\left[\int_{-\infty}^\infty\frac{(u-u')^i +a}{\left[b-(u-u')^2\right]^l}\right.
\nonumber\\
\times
\left.\frac{1}{1+u^{2n}}du\right]\frac{1}{1+u'^{2n}} du',
\label{int}
\end{eqnarray}

The integral between brackets can be solved using the residue theorem. First we notice that the zeros of the function $(u-u')^2-b$ are given by $u = u'\pm\sqrt{b}$, and in all cases, $b>0$. 
So we may avoid these zeros by adding an imaginary term $i\epsilon$ as $u' \rightarrow u' + i\epsilon$, $\epsilon>0$. Moreover, we may factorize the polynomial $1+u^{2n}$ as
\begin{equation}
u^{2n}+1 = \prod_{p=0}^{2n-1}(u-\psi_{n,p}),
\label{uu'}
\end{equation}
whose $2n$ roots are given by 
\begin{equation}
\psi_{n,p} = e^{i\frac{\pi}{2n}(1+2p)}.  
\label{roots}
\end{equation}
\begin{figure}[h!] 
\includegraphics[width=8.7cm]{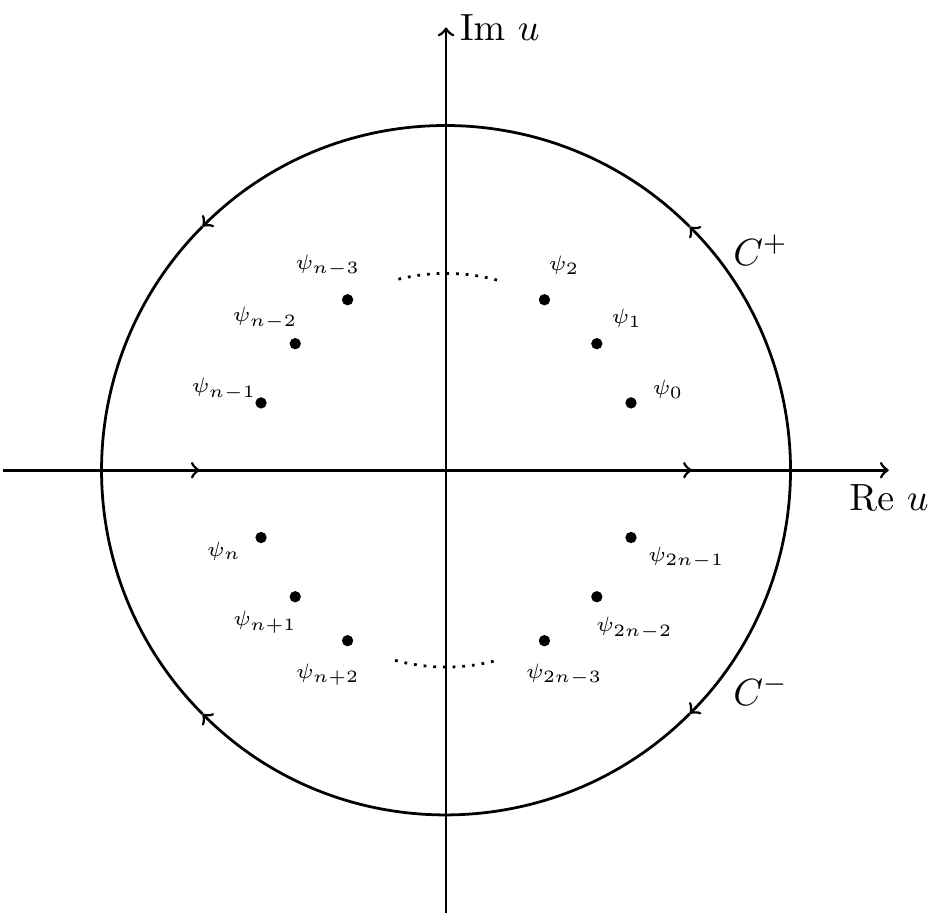}
\caption{Distribution of the roots of the polynomial $1+u^{2n}$ in the complex plane. Notice that there are $n$ roots in each half-plane. Here we have the shorthand notation $\psi_{n,p} \equiv \psi_p$.}
\label{contour}
\end{figure}
These $2n$ roots are distributed over a unit circle in the complex plane, as shown in Fig.~\ref{contour}. Notice that for each root in the upper half-plane there will be a mirror one in the lower half-plane. So by closing the contour in the lower half-plane ($C^-$) we will obtain exactly $n$ poles inside the contour, and using the residue theorem we find that
%
\begin{align}
&\int_{-\infty}^\infty\frac{(u-u')^i +a}{\left[b-(u-u')^2\right]^l}\frac{1}{1+u^{2n}}du 
\nonumber\\
&= \lim_{\epsilon \rightarrow 0}
\int_{C^-}\frac{(u-u')^i +a}{\left[b-(u-u'+i\epsilon)^2\right]^l}\frac{1}{\prod_{q=0}^{2n-1}(u-\psi_{n,q})}du 
\nonumber\\
&= -2\pi i \sum_{q=n}^{2n-1}\frac{(\psi_{n,q}-u')^i +a}{[b-(\psi_{n,q}-u')^2]^l}\frac{1}{\prod_{p\neq q}(\psi_{n,q}-\psi_{n,p})}.
\label{uint2}
\end{align}
%
Inserting this result into Eq. (\ref{int}), and using Eq. (\ref{uu'}) for the $u'$ variable, we get
\begin{eqnarray}
I_{\tau,ijkl}^{(n)}=-\left(\frac{\tau}{2}\right)^{2+i-2l}2\pi i \sum_{q=n}^{2n-1}\frac{1}{\prod_{q\neq p}(\psi_{n,q}-\psi_{n,p})}
\nonumber\\
\times\int_{-\infty}^\infty\frac{(\psi_{n,q}-u')^i +a}{[b-(\psi_{n,q}-u')^2]^l}\frac{1}{\prod_{r=0}^{2n-1}(u'-\psi_{n,r})} du'.
\end{eqnarray}

The remaining integral can be solved by following the same method used to solve the integral in the variable $u$, but now we close the contour in the upper half-plane ($C^+$).  Hence, 
\begin{eqnarray}
I_{\tau,ijkl}^{(n)}=4\pi^2\left(\frac{\tau}{2}\right)^{2+i-2l} \sum_{p=0}^{n-1}\sum_{q=n}^{2n-1}\frac{1}{\prod_{r\neq p}(\psi_{n,p}-\psi_{n,r})}
\nonumber\\
\times\frac{1}{\prod_{s\neq q}(\psi_{n,q}-\psi_{n,s})}\frac{(\psi_{n,q}-\psi_{n,p})^i +(2z)^j(2/\tau)^i}{[(2z)^k(2/\tau)^2-(\psi_{n,p}-\psi_{n,q})^2]^l}.
\label{result}
\end{eqnarray}

Looking at Eq. (\ref{roots}), we see that symmetries in $\psi_{n,p}$  allow a rather convenient simplification of the above result. Using the fact that
\begin{equation}
\frac{1}{\prod_{p\neq q}(\psi_{n,q}-\psi_{n,p})} = -\frac{\psi_{n,q}}{2n},
\end{equation}
we finally obtain
\begin{eqnarray}
I_{\tau,ijkl}^{(n)}=\frac{\pi^2}{n^2}\left(\frac{\tau}{2}\right)^{2+i-2l} \sum_{p=0}^{n-1}\sum_{q=n}^{2n-1}\psi_{n,p}\psi_{n,q}
\nonumber\\
\times 
\frac{(\psi_{n,q}-\psi_{n,p})^i +(2z)^j(2/\tau)^i}{[(2z)^k(2/\tau)^2-(\psi_{n,p}-\psi_{n,q})^2]^l}.
\label{result2}
\end{eqnarray}

By using this result to solve Eqs. (\ref{eqvx2}) and (\ref{eqvz2}), we find the dispersions given by Eqs.~(\ref{53}) and (\ref{53b}).

\end{document}